\documentclass[11pt,preprint,flushrt]{aastex}
\usepackage{amsmath}
\usepackage{amssymb}
\usepackage{latexsym}
\usepackage{cancel}
\usepackage{verbatim}
\usepackage{indentfirst}
\usepackage{color}
\usepackage{graphicx}

\begin{document}

\title{On-sky measurements of the transverse electric fields' effects in the Dark Energy Camera CCDs}
\author{A. A. Plazas$^{a\dagger}$, G. M. Bernstein$^b$, \& E. S. Sheldon$^a$}
\email{$^\dagger$aplazas@bnl.gov}
\affil{$^a$Department of Physics, Brookhaven National Laboratory, Bldg. 510, Upton, NY, 11792}
\affil{$^b$Department of Physics and Astronomy, University of Pennsylvania, Philadelphia, PA,  19104}

\begin{abstract}
Photo-generated charge in thick, back-illuminated, fully-depleted CCDs is transported by electric fields from the silicon substrate to the collecting well at the front gate of the CCDs. However, electric fields transverse to the surface of the CCD --with diverse origins such as doping gradients, guard rings around the imaging area of the sensor, and physical stresses on the silicon lattice-- displace this charge, effectively modifying the pixel area and producing noticeable signals in astrometric and photometric measurements. We use data from the science verification period of the Dark Energy Survey (DES) to characterize these effects in the Dark Energy Camera (DECam) CCDs. The transverse fields mainly manifest as concentric rings (\emph{tree rings}) and bright stripes near the boundaries of the detectors (\emph{edge distortions}) with relative amplitudes of about $1\%$ and $10\%$ in the flat-field images, respectively. Their nature as pixel size variations is confirmed by comparing their photometric and astrometric signatures. Using flat-field images from DECam, we derive templates in the five DES photometric bands ($grizY$) for the tree rings and the edge distortions as a function of their position in each DECam detector. {These templates can be directly incorporated into the derivation of photometric and astrometric solutions, helping to meet the DES photometric and astrometric requirements.} 
\end{abstract}

\section{Introduction}\label{sec:intro}

In the past decade, the development of thick, high-resistivity CCDs with high quantum efficiency (QE) at long wavelengths (near infrared) has been encouraged by increasing scientific interest in this part of the electromagnetic spectrum {\citep{holland2003,holland2007}}. Thick CCDs also offer other advantages over more conventional and thin CCDs by reducing fringing at long wavelengths. Thus, thick, fully-depleted CCDs have been chosen by several current and future astronomical surveys {and imagers} (\emph{e.g.}, the \emph{Dark Energy Survey}, DES\footnote{\url{www.darkenergysurvey.org}} {\citep{abbott2005}}, {the \emph{Pan-STARRS} survey \citep{kaiser2010}, the \emph{Hyper Suprime Camera} \citep{komiyama2010}}, and the \emph{Large Synoptic Survey Telescope}, LSST\footnote{\url{www.lsst.org}} {\citep{ivezic2008}}). {Despite their advantages, thick CCDs also exhibit undesirable characteristics that can leave signatures in the data, with consequences for shape, point spread function (PSF), astrometric, and photometric measurements \citep{stubbs2014,lupton2014,antilogus2014,jarvis2014}}.  

{Current and future experiments, such as DES and LSST, will produce large data sets, which will provide higher statistical precision. Thus, it has become} necessary to fully characterize and understand even the most subtle systematic effects in the methods and instrumentation {\citep{weinberg2013}}.  For instance, supernovae probes are limited by color and flux calibration, not statistical errors. In addition, weak gravitational lensing (WL) of large scale structure of the Universe (cosmic shear) requires measuring galaxy shapes and positions with high accuracy in order to exploit its full potential to constrain dark energy \citep{albrecht2006}.  

In this paper, we study the consequences of electric fields transverse to the surface of the CCDs of the Dark Energy Camera (DECam, {\cite{diehl2012,flaugher2012}; Flaugher et al. (in prep.)}), the imager built for the DES project. These transverse fields\footnote{Also referred to as ``lateral fields".} have an impact on astrometric and photometric measurements by shifting the location of the collected charge and modifying the effective pixel area. Charge relocation and pixel area variations contribute to the local variations in the response to uniform illumination (Pixel Response Non-Uniformity, PRNU), which are usually assumed to originate purely from sensitivity (or QE) differences in the pixels. Thus, naively dividing by the flat-field images could lead to systematic errors in the calibration steps during the data reduction process.

In the next Section, we provide a brief introduction to DECam in the context of the DES and illustrate some of the structures visible in flat-field images that do not correspond purely to pixel sensitivity differences. In Section $\S$3, we use flat-field images to derive templates of the amplitude of these effects as a function of their pixel location in the detectors. We then demonstrate, in Section $\S$4, how these templates can be incorporated into astrometric and photometric solutions to reduce residuals below the scientific requirements for DES. We conclude and summarize our results and their implications in Section $\S$5.

\section{The Dark Energy Survey and the Dark Energy Camera}\label{sec:intro-des}
The Dark Energy Survey is a multi band ($grizY$) photometric survey of 5000 sq deg of the Southern sky whose main scientific purpose is to constrain the dark energy equation of state, $w$. Four independent probes--weak lensing, galaxy cluster counts, baryon acoustic oscillations, and type Ia supernovae--will be combined under this single experiment.  A new 570 megapixel camera, the Dark Energy Camera, {was built at the Fermi National Accelerator Laboratory (Fermilab, 2008-2012),} and commissioned on the 4m Blanco telescope in the Cerro Tololo Interamerican Observatory (CTIO) in La Serena, Chile. After commissioning, {DES carried out a science verification process} (Nov. 2012--Feb 2013) to assess the quality of the images.

The focal plane of DECam is composed of 62 2k by 4k CCDs, plus 12 more 2k by 2k CCDs for focus and guiding. These devices are back-illuminated, fully-depleted (high-resistivity), thick ($\approx 250 \mu$m), n-type (p-channel) CCDs, {supplied by Teledyne DALSA Semiconductor} {and Lawrence Berkley National Laboratory \citep{holland2003}, and packaged \citep{derylo2006} and tested \citep{diehl2008,estrada2010} at Fermilab}. The pixel size of the devices is $15\ \mu$m, with a plate scale of about  $0''27$, generating a field of view of {about} $3$ deg$^2$.  

Detailed descriptions of DES, DECam, and the testing and characterization process of the CCDs can be found in {\citet{abbott2005,diehl2012,flaugher2012,diehl2008}; and \citet{estrada2010}}.

\subsection{Structures in flat-field images:  ``tree rings" (impurity gradients), edge distortions, and tape bumps (lattice stresses)}\label{sec:intro-flats}

\begin{figure}[ht]
\centering
\includegraphics[width=1.\textwidth]{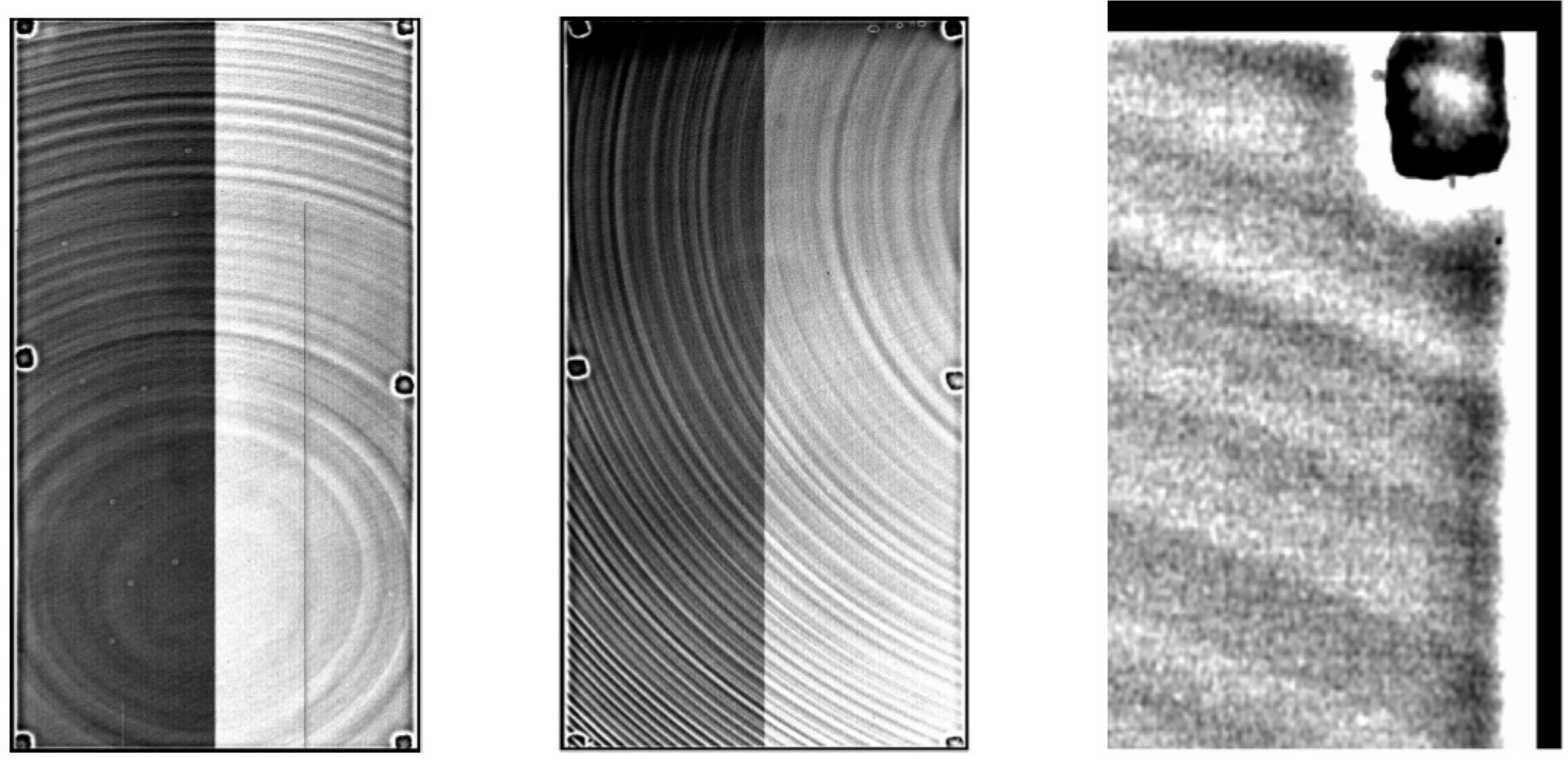}
\caption{Master dome flat images from two of the DECam CCDs in the $g$ band. Each image is normalized to 1, and the gray scale represents deviations of approximately $\pm 1\%$ from this value. Lighter shades represent enhanced brightness (the black strip surrounding the CCD is a 15 pixels-wide area, masked during the making of the master flats). The circular structures known as ``tree rings" can be seen in the first two images. {The tree rings are nearly symmetric about the axis of the silicon boules from which the CCD wafers were cut (there are four 2k by 4k CCDs per wafer), so the patterns on each CCD will depend on their relative position in the wafer}. The contrast difference between the left and right sides is due to gain differences in the two readout amplifiers of the detector. The last image shows the effect of the distorted electric field at the edge (bright stripes) and that of the lattice stresses due to double-sided tape (upper right corner).}
\label{plotone}
\end{figure}

Dome flats are taken daily {\citep{marshall2013}} as part of standard DES operations \citep{diehl2014} for the usual data reduction and calibration process.\footnote{{Throughout this paper, we refer to data processing for the Dark Energy Survey.}} A single, reduced master dome flat\footnote{A master dome flat is the median of the best dome-flat exposures of each night, reduced, and normalized to 1.} per CCD and for each photometric band is produced by one of the DES Data Management pipelines {\citep{mohr2008,desai2012}}. The most visually striking features in the flats are concentric circles, commonly known as ``tree rings'' (see Figure \ref{plotone}). They originate from circularly symmetric variations in impurity concentrations that  arise during the growth of single-crystal silicon from polysilicon (see Figure 5 in \citet{altmannshofer2003}).\footnote{High resistivity single crystal silicon is grown by means of the zone melting technique through the floating zone (FZ) method \citep{vonAmmon1984}. In the fabrication of DECam CCDs, the usual Czochralski process (CZ) method of single crystal growth is not used because the fused-silica crucible introduces large quantities of oxygen in the molten polysilicon, which limits the resistivity of the resulting single-crystal silicon to about 50 $\Omega\cdot$cm. Fully-depleted CCDs need to be fabricated on silicon with resistivity of about 5000 $\Omega\cdot$cm or more.} If the density of doping impurities is not uniform in the direction parallel to the back-side of the CCD (the $x$ direction), a difference in space charges will generate a transverse field $E_{\perp}$, perpendicular to the optical axis of the detector (the $y$ direction). A charge packet generated in the depletion region will be transported to the collecting region near the front gate of the CCD by a combination of this extra transverse field and the usual electric field in the device. 
This transverse field deflects the charge from its nominal path, {implying that the pixels in the device map to different areas on the incident side, some larger and some smaller. We say that the transverse fields change the \emph{effective area} of individual pixels in the detector, making some subtend more or less solid angle of sky than the average, and modifying the observed flux because the collected charge is conserved and no carriers are absorbed or created, just re-distributed between neighboring pixels.} 

Another important effect seen in the dome flats of the DECam devices is an increase in brightness near the four detector edges, likely due to the change in electric fields imposed by the boundary conditions and particular electronic structures (\emph{e.g}, front-gate electrodes, guard/floating rings) at the edge. These edge distortions are colloquially known in DES devices as ``glowing edges". A similar feature has been measured in LSST devices, but with an opposite sign (\emph{i.e.}, a reduction in brightness is seen closer to the edge) {\citep{oconnor2014}}. Therefore, we will refer to this effect more generally as the edge-distortion effect. In the DECam devices, this effect is measurable within about 100 pixels from the edge. Within 8-10 pixels from the edge, the relative counts are approximately 10$\%$ larger than the average.  

Finally, three small (about 50 by 50 pixels) deformations on each side of the CCD can also be seen. These features arise from physical stresses on the silicon lattice (that in turn distort the electric field lines) due to pieces of double-sided tape between the CCD and its mount.{\footnote{{More precisely, these deformations originate due to the difference between the coefficient of thermal expansion of the glue (Epotek 301-2) and the tape (Mylar), \citep{derylo2006}.} }} {To handle these tape bumps in DES, a flag for these pixels will be implemented to indicate that they are unreliable for high-precision analyses}. Therefore, we will discuss only the tree rings and edge distortions hereafter.

The tree rings and the edge distortions both originate during the manufacturing of the devices, while the tape bumps originate during the packaging process. They can be classified as \emph{fixed pattern} distortions, as opposed to other \emph{dynamic} effects of transverse fields such as pixel-to-pixel correlations in the collected charge that originate intensity-dependent PSFs \citep{rasmussen2014,antilogus2014}. These effects were already identified during the CCD packaging and testing at Fermilab \citep{derylo2006,diehl2008,estrada2010}. In particular, the edge distortions have been measured as astrometric displacements \citep{kuhlman2011}, and modeled in simulations \citep{holland2009}. Our measurements, however, are the first to demonstrate that the tree-rings effect is due to transverse electric fields. 

The amplitudes of the tree-ring, lattice-stress, and edge-distortion effects depend on the wavelength of the incoming signal, according to the absorption length in silicon. It is also expected that their amplitude will depend inversely on the value of the substrate voltage $V_{\text{sub}}$ applied to the back side of the CCD to reach overdepletion,\footnote{For an overdepleted CCD, the transit time of a carrier is proportional to $(V_{\text{sub}} - V_{\text{J}})^{-1}$, where  $V_{\text{J}}$ {is the average channel potential at the junction between the p-channel and n-type substrate \citep{holland2003}}. For an underdepleted device, the transit time is dominated by the time spent in the field-free region \citep{fairfield2006}.} although the SV data analyzed in this work were taken with $V_{\text{sub}}$ having a fixed nominal value of $40$V.

Flat-field images generally can be used to infer variations in the properties of pixels on a CCD. These variations may arise from intrinsic changes in pixel sensitivity (QE), or pixel area variations \citep{smith2008,kotov2011}. Therefore, it is not clear \emph{a priori} whether the effects mentioned above are due to one or the other. Since the main purpose of dividing the scientific images by a flat field frame is to account for the sensitivity differences across the pixel array, dividing by a flat field image that contains effects other than sensitivity variations will introduce systematic errors in the data. Thus, it is important to characterize the nature of these effects.  

Data taken with DECam during the DES science verification period show correlations between the tree ring patterns in the dome flats and astrometric displacements (see next Section and Figure \ref{plotfour})\footnote{The tape bumps show these astrometric signatures too, as can also be seen in Figure \ref{plotfour}}. This strongly supports the interpretation that they are due to charge redistribution between neighboring pixels caused by the presence of spurious electric fields transverse to the optical axis of the device, and not just quantum efficiency variations.

\section {Impact on astrometry and photometry}
\subsection {Star flats, and astrometric and photometric solutions}
By examining the residuals in the astrometric and photometric solutions, we can identify patterns that remain due to the effects of the transverse electric fields. We use sets of about 20 to 25 dithered exposures of star fields (per photometric band) {taken during SV operations}, in which each star is recorded at different parts of the detector.  From these exposures, we derive a \emph{star flat} image {\citep{manfroid1995,tucker2007}}, which is the ratio of the response of the instrument to focused light (light from stars) and its response to diffuse light (flat-field images).  

We divide the input images by the dome flats, and measure the instrumental magnitude for each star through large-aperture stellar photometry. A ``star-flat" photometric correction model is produced by varying the parameters of the model to minimize the disagreement between magnitudes of the multiple measurements of each star. This model is allowed to have low-order polynomial variations across each CCD, plus an exposure-dependent offset and linear slope across the full array.  Every measurement of each star has a residual of its star-flat-corrected magnitude to the mean magnitude of all measurements of that star.

In the astrometric models, we use algorithms that improve on the initial astrometric solution derived by the software {\tt SCAMP} \citep{bertin2006}. We solve for the parameters of the geometric distortion model (the map between pixel coordinates and sky coordinates, $\Omega(x,y)$\footnote{Also known as the World Coordinate System (WCS) solution.}) by forcing agreement between positions of a given star in different exposures, and then calculating the astrometric residuals with respect to the mean position of the star. 

After deriving the solutions, we record the astrometric and photometric residuals as a function of focal plane position. Over the five years of the full survey, the DES science requirements demand an astrometric accuracy $< 15$ mas for the centroids of objects in different exposures, and a relative photometry of $2 \%$ {\citep{annis2010}}.

\subsection{Photometric signatures}

\begin{figure}[tp] 
\centering
\includegraphics[width=0.75\textwidth, page=1]{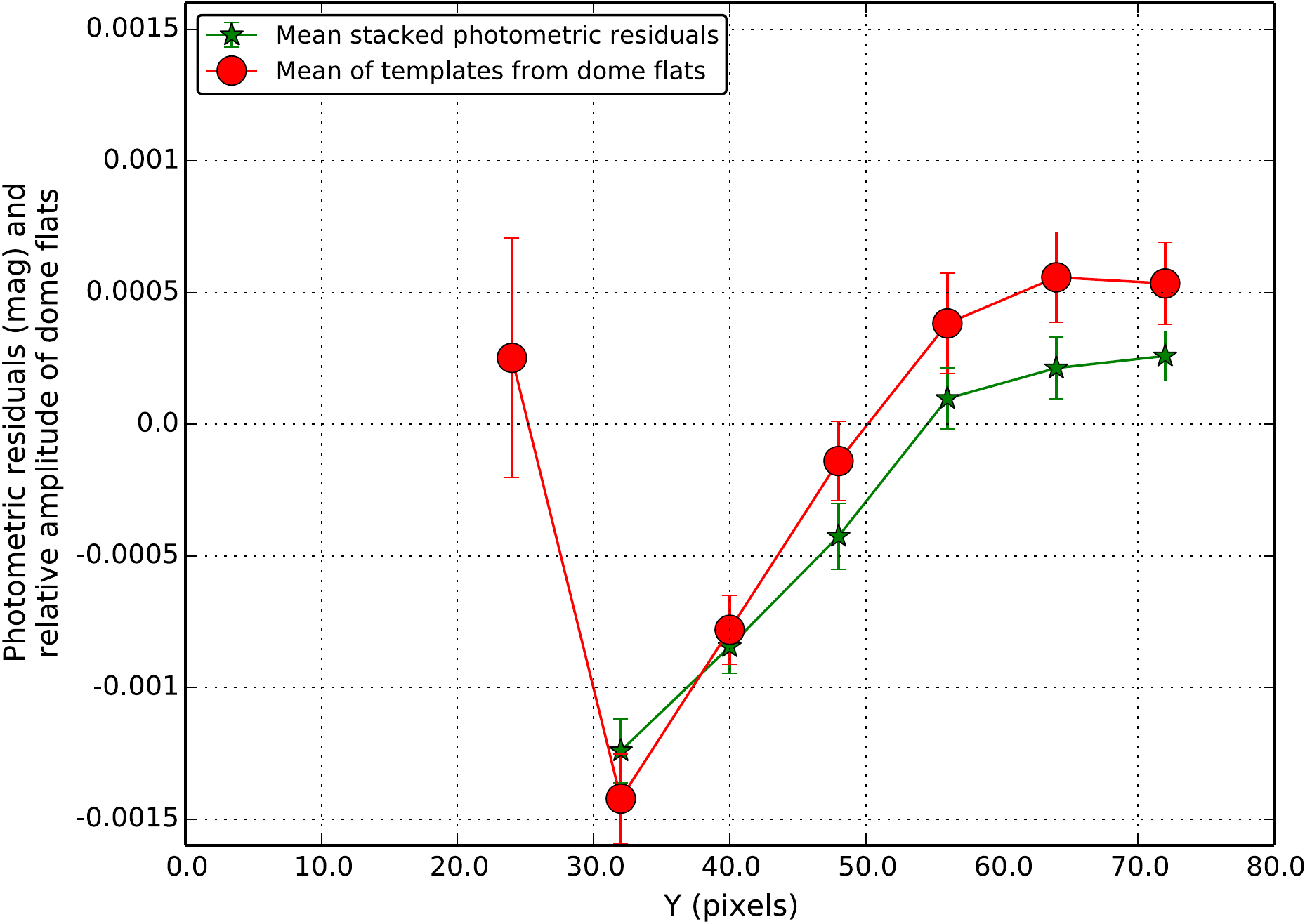}
\caption{Stacked photometric residuals (green star-shaped symbols) and stacked dome-flat signal (red dot-shaped symbols) as a function of distance of the bottom edge (relative to the CCD layout shown in Figure \ref{plotone}) of the detectors, binned every 8 pixels. {The distortions at the other three edges have similar shapes, but their amplitudes differ}. The stack is over all CCDs of the focal plane and all photometric bands. The input images for the photometric solution were divided by the dome-flat images.}
\label{plottwo}
\end{figure}
For each CCD and filter, we bin the photometric residuals by their distance from each of the four edges. We then compare the residuals to binned signal from the flat-field data, including a multiplicative factor to match the two amplitudes, when necessary. In most cases, this factor deviates from unity, although on average both signals are correlated with each other, as seen in the example of Figure \ref{plottwo}.

\begin{figure}[tbp] 
\centering
\includegraphics[width=0.8\textwidth]{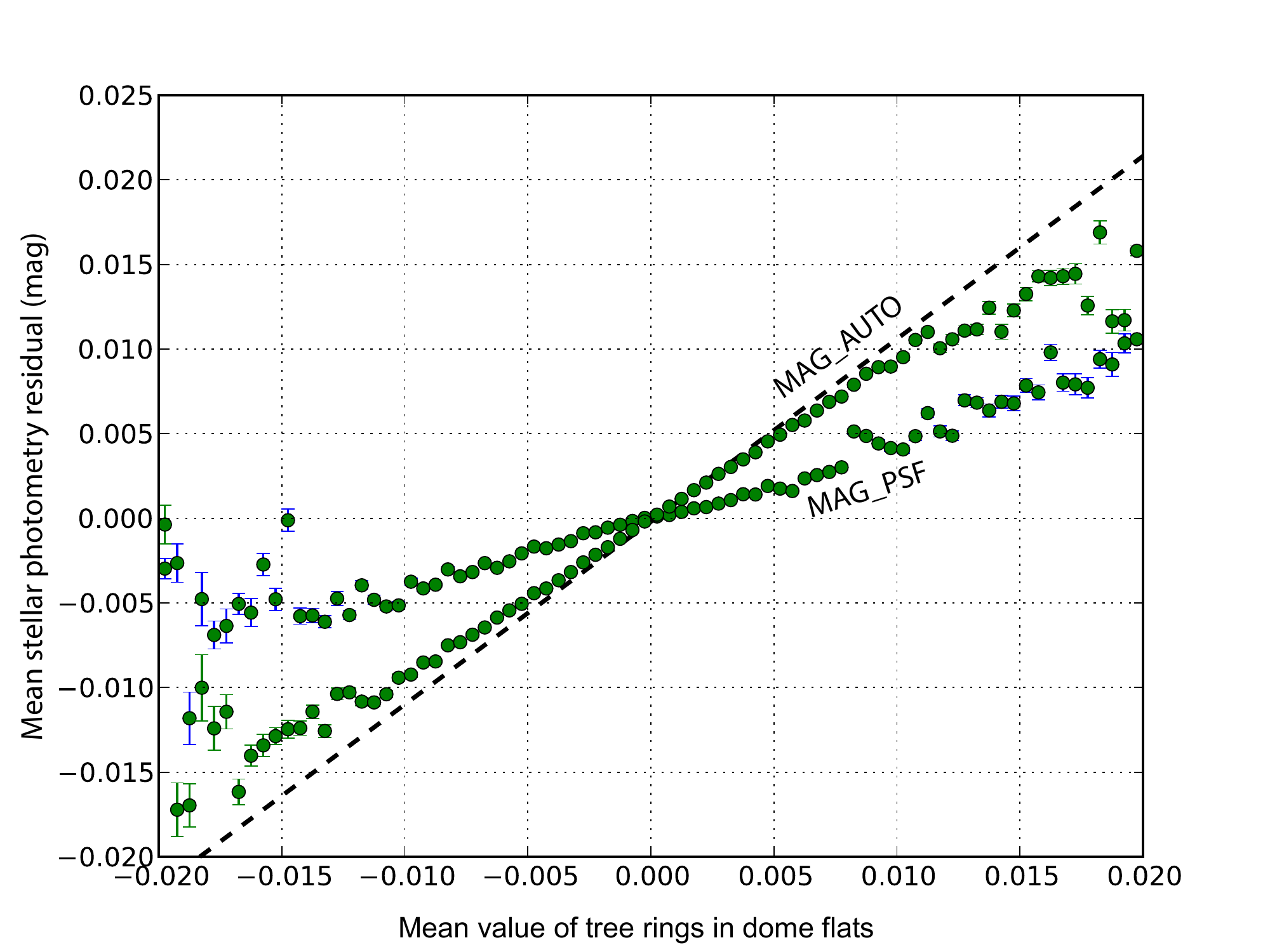}
\caption{Mean amplitude of tree rings as measured on the dome flats compared to means of photometric residuals in bins of tree ring values from $i$-band DES data. The input images for the photometric solution were divided by the dome-flat images. The aperture of each object was adjusted and measured with the {{\tt SExtractor}} \citep{bertin1996} parameter {{\tt MAG\_AUTO}}. On average, the data trace the dotted line, which has a slope of $2.5\ln(10) \approx 1.085$, indicating a positive correlation between the two quantities. The damping at large tree-ring values is likely due the smoothing of the sharpest rings due to the finite size of the stellar images. When using fitting photometry ({{\tt SExtractor}} parameter {{\tt MAG\_PSF}}), there is an offset of $\approx2$x because both the flux and the size of the stars are different in the rings, so the PSF fitter (not knowing about the size variation) calculates an erroneous value for the total flux.}
\label{plotthree}
\end{figure}

For the signal produced by the tree rings, the binned photometric residuals trace the tree rings signal as measured in the dome flat images, as seen in Figure \ref{plotthree}.

In both cases, the residuals are calculated after dividing by the dome-flat images, and the photometric model for these data does not include terms to track the edge distortions and the tree rings. Hence, the correlations between the dome-flat signals and the photometric residuals indicate that the tree-rings and edge distortion features are not tracing pixel-sensitivity variations. 

\subsection{Astrometric signatures} 
\begin{figure}[tbp] 
\centering
\includegraphics[width=0.85\textwidth, page=1]{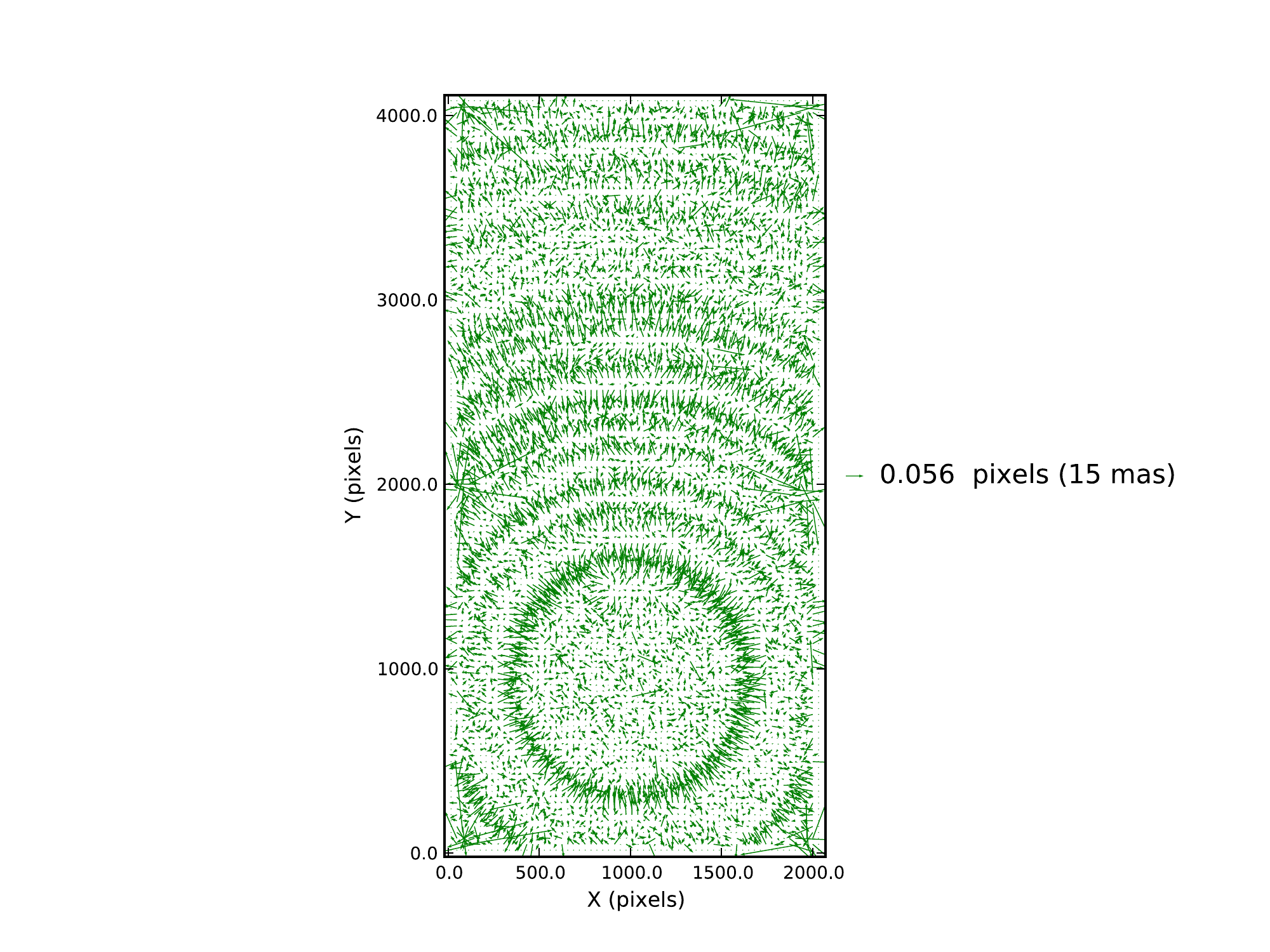}
\caption{Astrometric residuals in the $x$ and $y$ directions plotted as a vector field for a particular DECam device. The input images for the astrometric solution were divided by the dome-flat images. The median magnitude across the whole chip is $5.65$ mas, with an rms of $5.35$ mas. Large residuals can be seen at the positions of the tree rings and the tape bumps (compare the dome flat images shown in Figure \ref{plotone}). The astrometric model did not contain terms to account for the edge distortions either, but their imprint on the residuals are partially hidden by the masking of 30 pixels at the edges when calculating the solution.}
\label{plotfour}
\end{figure}

As with the photometric residuals, we plotted the astrometric residuals as a function of their position in the CCDs, for each detector and each band. The astrometric model did not contain terms to fit for the tree rings or the edge distortions.\footnote{Or a term to take into account the tape bumps, but these will be flagged.} In Figure \ref{plotfour} we show the vector field of astrometric residuals as a function of CCD position for a particular device, stacked over all exposures in all photometric bands. Correlations are apparent between the tree-ring patterns and the tape bumps from the dome flat images. The edge distortions also leave large astrometric residuals, but they are partially hidden by the masking of 30 pixels at the detector edges in the calculation of the astrometric solution. 

\begin{figure}[tbp] 
\centering
\includegraphics[width=1.\textwidth]{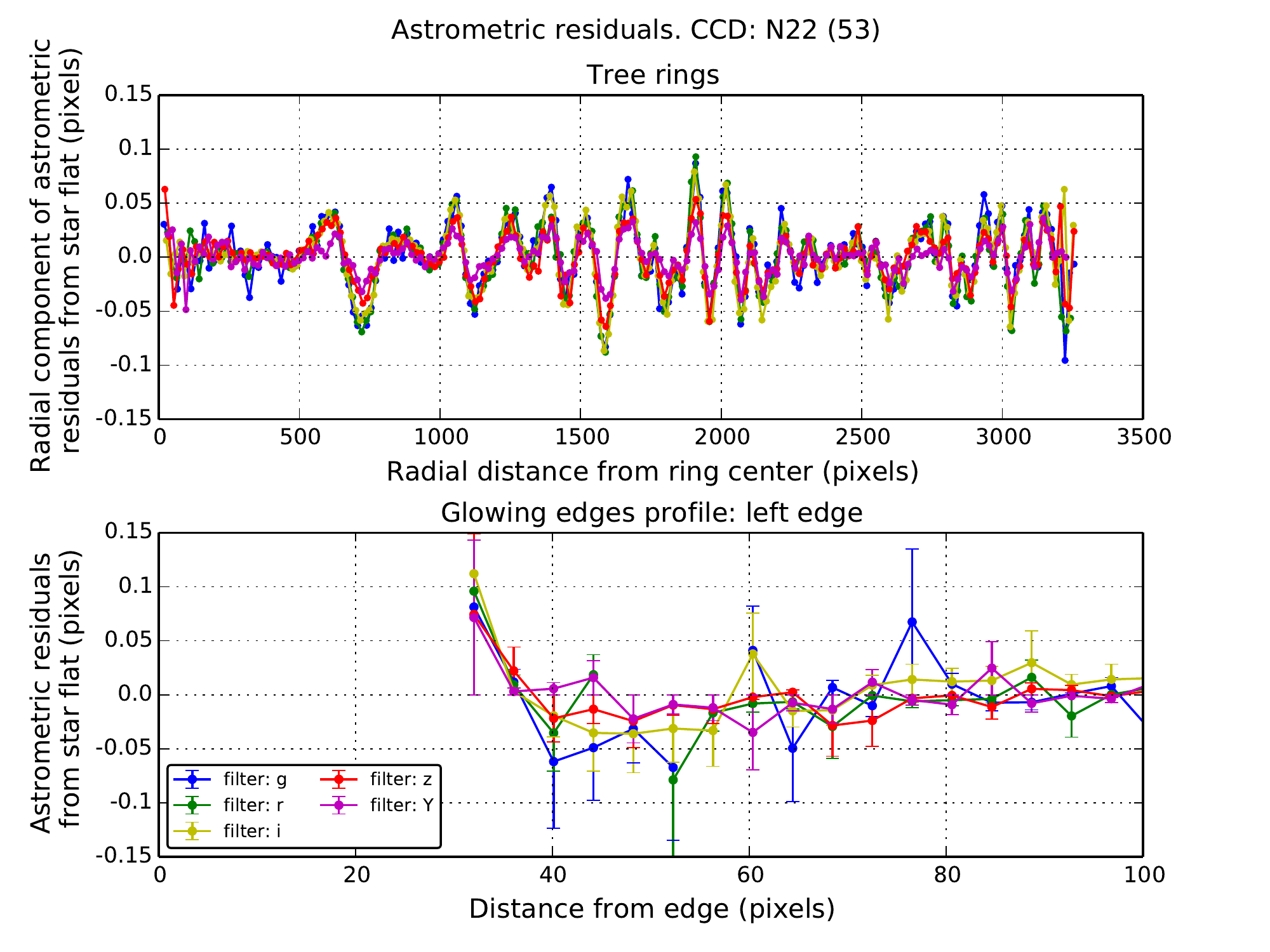}
\caption{Tree-ring (upper panel) and edge-distortion (bottom panel) astrometric profiles as a function of distance from the center of the rings and the left edge, respectively, for a particular DECam device, and for each of the five DES photometric bands. The tree-ring and edge-distortion signals were averaged in bins of 16 and 4 pixels, respectively. The average scaling factors (with respect to the $g$ band) for the tree-ring profiles of all DEcam CCDs are shown in the appendix (Figure \ref{ploteight}). The input images for the astrometric solution were divided by the dome-flat images. The imprint on the residuals at the edge are partially hidden by the masking of 30 pixels at the edges when calculating the astrometric solution.}
\label{plotfive}
\end{figure}

After identifying the center of the rings in each device,{\footnote{{It is possible to infer the pixel coordinates of the common center of the tree rings in each CCD by knowing the relative position of each one with respect to the center of the silicon wafer from where it was cut. However, we found that the center coordinates obtained in this way were not accurate, and therefore we fitted for them by identifying points on the circumferences of several rings.}}} we {azimuthally average the signal in order to} create a radial profile of the tree rings' signal as a function of distance from that center. In an analogous way, we create profiles of the astrometric signatures as a function of the distance to each edge {by averaging the data along columns} (Figure \ref{plotfive}). The amplitude of the astrometric signal in the case of the rings is approximately $0.05$ pixels ($\approx13$ mas) to $0.1$ pixels ($\approx 26$ mas); for the edge distortion, the amplitude is also approximately $0.05$-$0.1$ pixels.\footnote{The relative amplitude of the edge distortion can be of $10\%$ or more as the distance to the edges diminishes. However, the master dome-flats we used in our analysis have a masked region of 15 pixels from the edge, and the astrometric and photometric solutions mask regions of 30 pixels within the CCDs boundaries.}  We also note in Figure \ref{plotfive} that both effects depend on wavelength.

\section{Improving the astrometric and photometric solutions using templates from dome flats}
\label{sec:templates}
\begin{figure}[tbp]
\centering
\includegraphics[width=1.\textwidth]{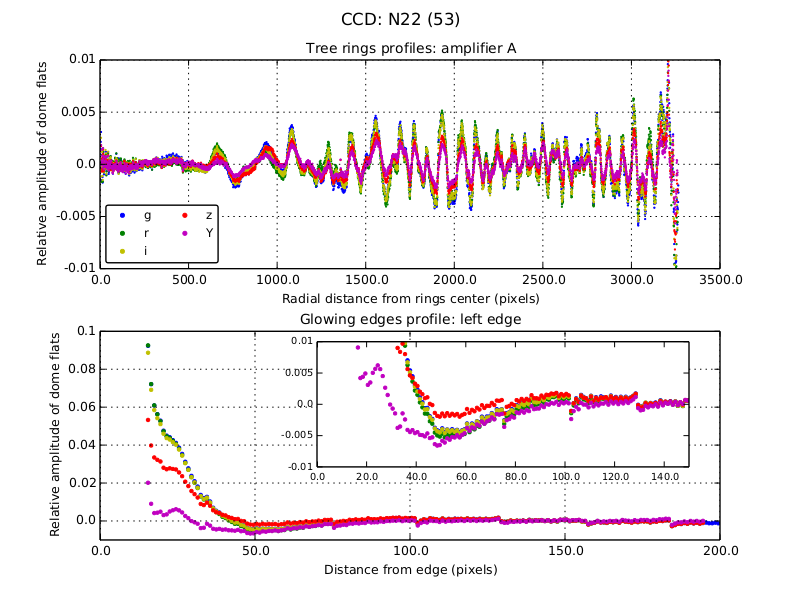}
\caption{Profiles of the tree rings and the left edge at different bands from dome flats for one of the DECam detectors. A region of 15 pixels within the edge is masked in the dome-flat images. The tree-ring profiles (upper panel) differ from each other by scaling factors ($g>r>i>z>Y$) at all distances from the center of the rings, whereas the profiles from the edge distortions (lower panel) exhibit a more complex wavelength dependence. The average scaling factors (with respect to the $g$ band) for the tree-ring profiles of all DEcam CCDs are shown in the appendix (Figure \ref{ploteight}).}
\label{plotsix}
\end{figure}

We would like to characterize the structures in the photometric and astrometric residuals by using the information from the high signal-to-noise (S/N) ratio dome-flat images (compared to the signal obtained from the finite number of stars in the star-flat images). For that purpose, we measure templates of the edge distortion as a function of the distance from each of the four edges.  We assume that any actual QE variations near the edges are suppressed by averaging rows (columns) in regions that do not include the tape bumps. For tree rings, we identify their centers in each one of the DECam devices and bin their amplitudes in the flats as a function of distances to these centers. As with the edges, we assume that QE variations as a function of distance from the ring center are gradual and removed by applying a high-pass filter to the measured radial profile.  We do this for each one of the five DES photometric bands ($grizY$). The wavelength dependence is more apparent than in the astrometric residuals due to the increased S/N ratio in the dome-flat images (see Figure \ref{plotsix}, compared to Figure \ref{plotfive}).  

In general, we would expect to measure a larger amplitude towards the blue part of the spectrum because the absorption length function of light in silicon implies that photons with shorter wavelength will interact, on average, closer to the back, incident side of the CCD. When a photon of wavelength $\lambda$ enters a back-illuminated CCD, it is absorbed in the silicon substrate at an average depth that depends on $\lambda$ (at $\lambda \approx$1060 nm silicon is practically transparent to incoming radiation). In an n-type detector, a hole is produced and carried to the collecting potential well in the buried p-type channel nearby the polysilicon gate electrodes by an electric field $E_{\parallel}(y).$\footnote{The $y$ direction is taken perpendicular to the CCD surface, with the origin at the rear window at the back of the CCD. The $x$ direction is parallel to the surface, forming a right-handed reference system with the $y$ axis.} This field has a {minimum magnitude} at the window and peaks at the p-n junction, where the holes are collected. Thus, photo-generated holes from blue light would be subject to the effect of transverse electric fields by more than for redder light, displacing the centroid of the charge packet. 

The ring radial profiles seem to be consistent with this model (Figure \ref{plotsix}), and each profile differs from the other only by a multiplicative factor. It is possible to produce estimates of the numerical value of these factors by positing a functional model for the $y$ dependence of the transverse and main electric fields, knowing the spectral energy distribution of the light sources (LEDs in the case of the flat-field images, and stars in the case of the star-flat images), and the probability density distribution of the interaction of photons with silicon. In Appendix A, we approximate these functions and compute values that qualitatively agree with the values measured (averaged over all the DECam detectors) using the profiles from the flat- and star-field images, for the tree-ring profiles. 
In the edge distortions, the wavelength dependence is not a simple scaling at all distances from the edge, suggesting that a more complicated electric field model might be needed to account for the observations at the edges (see below).

The measured tree rings and edge templates from the dome flats can be incorporated in the photometric model described above to improve the photometric solution as functional terms  with varying amplitudes. 
 
\begin{figure}[tbp]
\centering
\includegraphics[width=1.\textwidth]{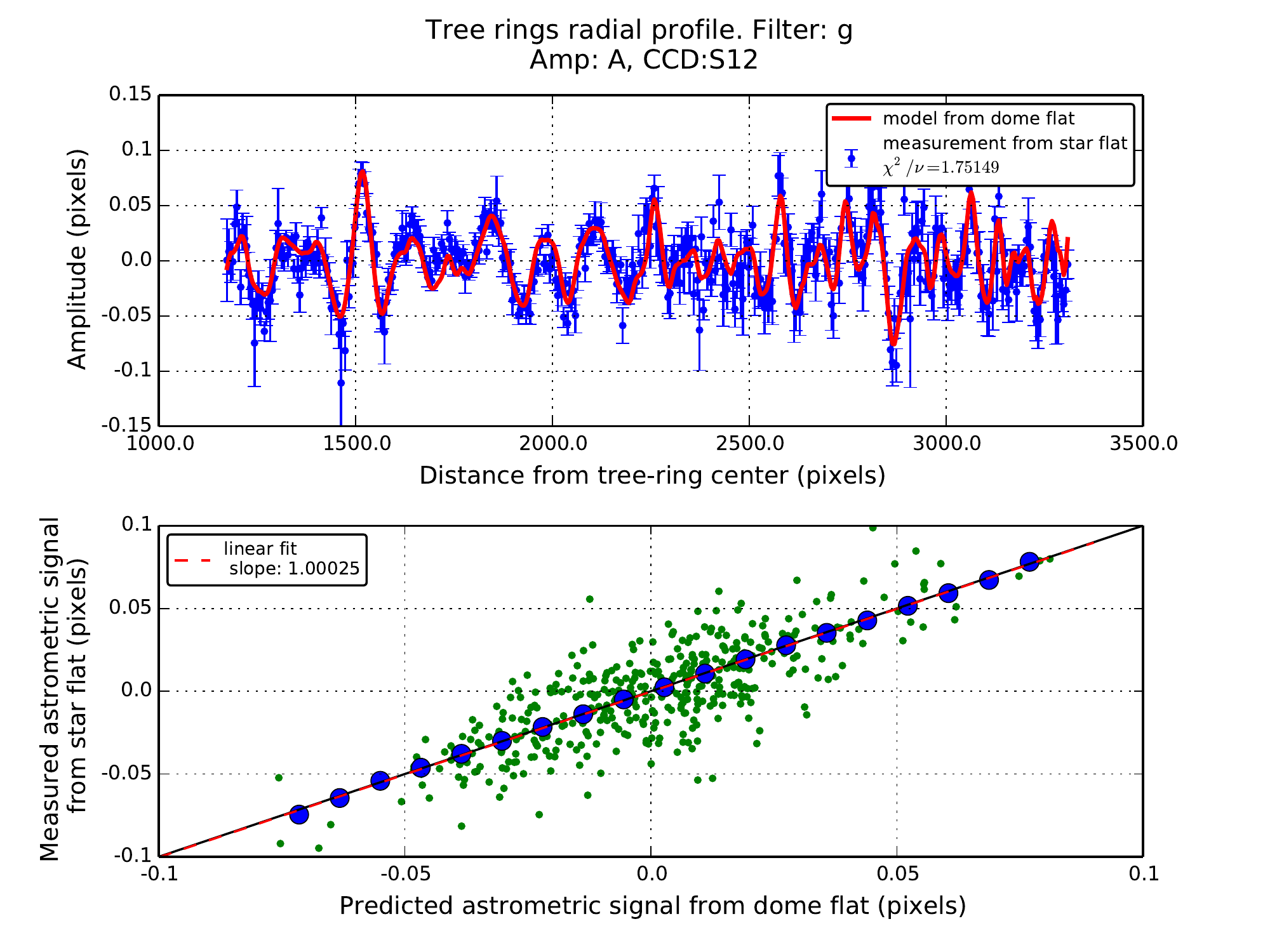}
\caption{Measured astrometric residuals, binned as a function of radial distance from the tree rings' center for a particular detector in the $g$ band. The prediction derived from the photometric signal of the dome flats  correlates with the data. The fitted function can be incorporated into the astrometric model to reduce the remaining residuals. }
\label{plotseven}
\end{figure}

To generate tree-ring templates for the astrometric solutions, we would like to utilize the photometric signal in the dome flats again. To show how the signal in the dome flats and the signal measured by the astrometric residual map are related to each other, we begin by pointing out that the number of photons per unit area in the pixels is proportional to the intensity of the incident light, which is uniform on small scales in a flat-field image. Photon number conservation therefore translates as sky-area conservation in the flat-field signals. In the tree rings, the astrometric errors can be modeled as a function, {$f(r)$}, of radius from the ring center, representing the displacement of photons hitting at position $r$ to photons detected at $r' \approx r + f(r)$. The dome-flat images include the effect of the variable pixel area across the field, and therefore measure their effective area, $A'$. The oscillatory tree-ring pattern measured in the dome flats can be written as a radial function $w(r)$, which is the azimuthally averaged tree-ring signal from the dome flats  (normalized to 1 and high-pass filtered). It is related to the Jacobian determinant of the coordinate transformation between this effective area and the unperturbed area $A$ by:
\begin{equation}
1 + w(r) = \left |\frac{dA'}{dA} \right |
\label{eq:jacobian}
\end{equation}
From Equation (\ref{eq:jacobian}), we can obtain an expression for the function $f(r)$ as a function of $w(r)$: 
\begin{equation}
1 + w(r) = \left |\frac{r' dr' d\theta'}{r dr d\theta} \right | 
\approx  \left| \frac{(r + f)(dr + dr \partial_r f)}{r dr} \right| \end{equation}
Thus, to first order in $f(r)$, we have: 
\begin{equation}
\label{eq:prediction}
{f(r) = \frac{1}{r} \int_0^r r' w(r') dr'}
\end{equation}
By using Equation (\ref{eq:prediction}) we can use the ring patterns measured from the dome flats to predict the radial component\footnote{The azimuthal component is consistent with zero.} of the astrometric error induced by the tree rings. The predicted astrometric function $f(r)$ (after removing large-scale modes introduced by the integration process) correlates with the measured radial component of the astrometric residuals from the star flats, as can be seen in Figure \ref{plotseven}. The astrometric predictions for each DECam CCD in each band can now be tabulated and incorporated into the astrometric model (by including them in the geometric distortion map between pixel and sky coordinates, $\Omega(x,y)$).

For the glowing edges, an analogous equation in cartesian coordinates should apply ($f(x) = \int w(x')dx'$) if the transverse electric fields behave in a similar way to the ones giving rise to the tree-ring profiles. However, we find disagreement between the measured astrometric distortion and the predicted function from the dome-flat images.  We can still use templates measured directly from the star-flat images in the astrometric solution, stacking several CCDs to increase the S/N ratio. We expect the profiles to be similar for a particular edge at a given band across all the devices,\footnote{At least in those CCDs that share similar fabrication characteristics, such as originating from the same silicon wafer.} unlike the radial profiles of the tree rings.  
 
These measurements, along with the non-monotonic wavelength dependence found in the flat field data (Figure \ref{plotsix}), are indications that other physical mechanism(s) might be at work at the edges of the detectors, in addition to pixel-size variations. Boundary conditions, in addition to electrodes at the frontside of the CCD and guard/floating rings surrounding the device, are likely to generate the electric field distortions at the edges. There are also holes from the non-imaging periphery of the CCD that drift to the active area that contribute to the measured signal.\footnote {Paul O'Connor, private communication.} \citet{holland2009} perform simulations of the electric fields near the edges of similar devices, illustrating the field deformations as a function of depth from the back window and distance from the edge, although their calculations assume boundary conditions that likely under-estimate the magnitude of the field distortion. Likewise, \citet{kuhlman2011} use x-ray-beams to measure the centroid shifts of a spot near the edge of the imager. In both cases, the calculations extend to 20 pixels away from the edge, while our measurements from the dome-flat (star-flat) images range from 15 pixels (30 pixels) to approximately 100 pixels from the edge. 

An appropriate model of the transverse electric fields at the edges would enable us to calculate the astrometric and photometric shifts as a function of wavelength and position from the edge. This would allow us to identify locations where the shifts are large enough to render the data unusable, according to the particular survey requirements. Chromatic gradients in the astrometric and photometric shifts (\emph{i.e.}, different shifts for ``blue" and ``red" photons) can also propagate into position and shape biases when averaged over different sources with particular spectral energy distributions (such as different star types, and different types of galaxies at different redshifts), in an analogous way to the effects caused by differential atmospheric refraction \citep{plazas2012}. {Efforts to simulate, model, and characterize transverse electric fields in CCDs have recently been pursued by \citet{holland2014} and \citet{rasmussen2014} }.

\section {Summary and Conclusion}\label{sec:end}

We analyzed the effects of electric fields transverse to the surface of the Dark Energy Camera CCDs on astrophysical measurements (photometry and astrometry). In particular, we studied the electric fields generated by circularly symmetric impurity gradients in the silicon wafer (giving origin to structures visually similar to tree rings), and distorted fields at the four edges of the detectors (``edge-distortion" effect). These structures are visible in the dome flat images used by the data calibration pipelines. 

We used data from the science verification period of the Dark Energy Survey to show how these instrumental features imprint a residual signal on astrometric measurements, indicating that they are consequence of charge redistribution among the different pixels of the CCD. The effective pixel area is modified too, leaving residual correlations in photometric measurements. {The measured photometric residuals left by these structures are about $0.5\%$  (Figure \ref{plotsix}), which is less than the required $2\%$ RMS of photometric precision for DES. However, taking into account these photometric signatures will contribute to achieve the DES goal of $1\%$ RMS photometric precision \citep{annis2010}. The astrometric residuals, on the other hand, have an amplitude about $13-26$ mas (Figure \ref{plotseven}), which will consume most of the relative DES astrometric error budget ($15$ mas) if left uncorrected}.   

We derived templates of the tree-ring and edge-distortion effects as a function of position in the focal plane, for each one of the DES photometric bands. These functions can be used in the modeling of the astrometric and photometric solutions for DES data. The measured photometric and astrometric shifts in different wavelengths from the tree-ring profiles are consistent with a static electric field that changes the pixel footprints on the sky, allowing the derivation of astrometric templates from the dome-flat data with higher S/N ratio. A similar comparison shows that the behavior at the edges is a combination of other physical effects in addition to pixel size variations.

Our analysis shows that the flat-fielding of raw images (as it is traditionally done to reduce astronomical data) can introduce systematic biases due to the presence of structures that do not purely trace the sensitivity variations of the pixels. This will have to be considered during the reduction and/or analysis of data from other wide-field cameras whose detectors also present tree rings and edge distortion effects (such as the LSST CCDs).

\section*{Acknowledgments}

We thank D. DePoy, H. T. Diehl, B. Flaugher, S. Holland, M. Jarvis, I. Kotov, T. Li, M. May, A. Nomerotski, P. O'Connor, A. Rassmusen, and W. Wester for useful comments and discussions.  
GMB is supported by Department of Energy grant DE- SC0007901 and from National Science Foundation grant AST-0908027.
AAP and ESS are supported by DOE grant DE-AC02-98CH10886.

{This paper has gone through internal review by the DES collaboration.} 

{We are grateful for the extraordinary contributions of our CTIO colleagues and the DES Camera, Commissioning, and Science Verification teams in achieving the excellent instrument and telescope conditions that have made this work possible.
The success of this project also relies critically on the expertise and dedication of the DES Data Management organization.}

Funding for the DES Projects has been provided by the U.S. Department of Energy, the U.S. National Science Foundation, the Ministry of Science and Education of Spain, the Science and Technology Facilities Council of the United Kingdom, the Higher Education Funding Council for England, the National Center for Supercomputing Applications at the University of Illinois at Urbana-Champaign, the Kavli Institute of Cosmological Physics at the University of Chicago, Financiadora de Estudos e Projetos, Funda\c{c}\~ao Carlos Chagas Filho de Amparo \`a Pesquisa do Estado do Rio de Janeiro, Conselho Nacional de Desenvolvimento Cient\'ifico e Tecnol\'ogico and the Minist\'erio da  Ci\^encia e Tecnologia, the Deutsche Forschungsgemeinschaft and the Collaborating Institutions in the Dark Energy  Survey.  \\

The Collaborating Institutions are Argonne National Laboratories, the University of California at Santa Cruz, the University of Cambridge, Centro de Investigaciones Energ\'eticas, Medioambientales y Tecnol\'ogicas-Madrid, the  University of Chicago, University College London, the DES-Brazil Consortium, the Eidgen\"ossische Technische  Hochschule (ETH) Z\"urich, Fermi National Accelerator Laboratory, the University of Edinburgh, the University of  Illinois at Urbana-Champaign, the Institut de Ciencies de l'Espai (IEEC/CSIC), the Institut de Fisica d'Altes Energies, the Lawrence Berkeley National Laboratory, the Ludwig-Maximilians Universit\"at and the associated Excellence Cluster Universe, the University of Michigan, the National Optical Astronomy Observatory, the University of Nottingham, the Ohio State University, the University of Pennsylvania, the University of Portsmouth, SLAC National Laboratory, Stanford University, the University of Sussex, and Texas A\&M University.

This research has made use of NASA's Astrophysics Data System.

\appendix
\section{Appendix A: Estimate of the relative amplitude of the tree-ring profiles in different bands}
\subsection{Analytic model.}

It is possible to estimate the relative amplitude of the tree-rings structures as a function of wavelength. For carriers created at a distance $y$ from the back of the CCD and letting $G(y)$ denote the astrometric/photometric displacement due to the tree rings, the average amplitude of the signal, $I_{T}(G(y))$, will be given by:   
\begin{equation}
\begin{split}
I_{\text{T}} (G(y)) = \frac{  \int_{\lambda_{\text{min}}}^{\lambda_{\text{max}}}d\lambda  \int_0^d dy \  \lambda T(\lambda) S_{\lambda}(\lambda)  f(y, \lambda) \  G(y)     +    \int_{\lambda_{\text{min}}}^{\lambda_{\text{max}}} d\lambda  \int_d^{2d} dy \ \lambda  T(\lambda) S_{\lambda}(\lambda)  f(y, \lambda) \ G(2d - y) } { \int_{\lambda_{\text{min}}}^{\lambda_{\text{max}}} d\lambda  \int_0^{2d} dy \ \lambda T(\lambda) S_{\lambda}(\lambda) f(y, \lambda) } 
\end{split}
\label{amplitude}
\end{equation}
In Equation (\ref{amplitude}), $T(\lambda)$ is the instrument response at a given photometric band (including terms such as the QE of the detectors, the optical elements in the telescope, and the atmospheric transmission\footnote{The atmospheric transmission is only included when calculating the mean astrometric shifts.}), $S_{\lambda}(\lambda)$ is the spectral energy distribution (SED) of the source, and $f(y, \lambda)$ is the absorption probability density of photons in the CCD. The second term in Equation (\ref{amplitude}) accounts for light that is reflected back from the front side of the device. The origin of the coordinate system is taken at the rear conductive window of the CCD, and $d$ represents the thickness of the detector ($d$ $\approx$ 250 $\mu$m for DECam CCDs). The extra factors of $\lambda$ account for the fact that the detectors are photon-counting devices.   

The distribution of absorbed photons at locations greater than $y$ is: 
\begin{equation}
P ( Y > y) = F(y) =  \frac{1 - \exp({-y\alpha(\lambda)})}{ 1 - \exp({-2d\alpha(\lambda)})}
\label{absorbed_photons}
\end{equation}
The denominator in Equation (\ref{absorbed_photons}) accounts for the photons that are never absorbed. 
Thus, the probability density of photons being absorbed between a location $y$ and $y + dy$ (denoted by $f(y, \lambda)$ in Equation (\ref{amplitude})) is given by: 
\begin{equation}
\frac{dF}{dy} \equiv f(y, \lambda) = \frac{\alpha \exp({-y\alpha(\lambda)})}{1- \exp({-2d\alpha(\lambda)})}
\end{equation}
$\alpha(\lambda)$ is the silicon absorption coefficient, given in inverse units of length. This function also depends on temperature. 

The astrometric shift caused by the transverse field $E_{\perp}(y)$ is denoted by $\Delta X_{\perp}(y)$, whereas the signature in the flat fields is given by the spatial derivative of this expression along the transverse axis $x$, $\partial_{x}\Delta X_{\perp}$ (see Equation (\ref{eq:prediction})). Therefore, to calculate the amplitude of the tree-rings profiles in different bands as measured in the dome-flat images we must set the functional $G(y)$ to the latter. In the same way, to estimate the amplitude of the astrometric shift caused by the tree rings, $G(y)$ should be set to $\Delta X_{\perp} (y)$. 

Assuming that the carriers' drift velocity is proportional to the electric field in the device ($v = \mu E$, where $\mu$ is the holes' mobility), we can write an expression for $\Delta X_{\perp}(y)$ in terms of the main drift field, $E_{\parallel} (y)$, and the perturbative transverse field, $E_{\perp}(y)$:
\begin{equation}
\Delta X_{\perp} (y) =  \int_y^{d} dt \ v_x(y')  =  \int_y^{d} dy' \ \frac{E_{\perp}(y')}{E_{\parallel}(y')}  
\label{delta_x}
\end{equation}

To integrate Equation (\ref{delta_x}), we need a model of the transverse and parallel fields. We want to calculate the relative wavelength scaling of the amplitude of the tree-ring profiles, which depends on the absorption length in the silicon substrate along the $y$ direction. Therefore, we will assume that it is possible to factor out the $x$ dependence of the transverse and parallel fields, and provide models only for $E_{\parallel} (y)$ and $E_{\perp}(y)$. When calculating the relative amplitude of the rings with respect to another band (e.g., the $g$ band), the $x$-dependent functions will cancel out. 

$E_{\parallel} (y)$ has three contributions: a constant term due to the fact that the pixel acts like a capacitor, a constant term due to space charges in the buried p-channel, and a linear term due to space charges in the n-type substrate. By solving the one-dimensional Poisson equation in a semiconductor with appropriate boundary conditions, $E_{\parallel} (y)$ can be written as a linear function of distance,  $E_{\parallel}(y) = -a\left( 1 + (b/a)(y/d -1)\right)$, with $a\equiv V_{\text{app}}/d + V_{\text{dep}}/d$ and $b\equiv qN_{\text{D}}d/\epsilon_{\text{Si}}$. $V_{\text{app}}$ is the voltage drop across the drift region, and it is equal to the difference between $V_{\text{sub}}$ and $V_{\text{J}}$ (the bias voltage applied to (over)deplete the device and {the average channel potential at the junction between the p-channel and n-type substrate}, respectively). The minimum voltage necessary to fully deplete the CCD is denoted by $V_{\text{dep}}=qN_{\text{D}}d^2/2\epsilon_{\text{Si}}$, where $q$ is the electron charge, $N_{\text{D}}$ is the substrate doping (donors in n-type silicon), and $\epsilon_{\text{Si}}$ is the permittivity of silicon \citep{holland2003}.\footnote{Note that the expression for $E_{\parallel}(y)$ that appears in \citet{holland2003} assumes that the origin is at the silicon-SiO$_2$ interface, at the front of the CCD. Our equations assume that the origin lies at the back side of the CCD.} 

Transverse differences in the space charges, or gradients in the direction perpendicular to the optical axis (the $x$ direction), give rise to $E_{\perp} (y)$. Boundary conditions imply that  $E_{\perp} (y)$ must vanish at the positions $y=0$ and $y=d$. In addition, simulations show that fields of this type peak roughly at $y=0.5d$ \citep{kotov2006}. Thus, we assume that the $y$ dependence of this field is of the structural form  $E_{\perp}(y) = A_{\perp} (y/d) (1 - y/d)$, with $A_{\perp}$ a constant to give dimensional consistency to this expression.

\begin{figure}[tbp]
\centering
\includegraphics[width=1.\textwidth]{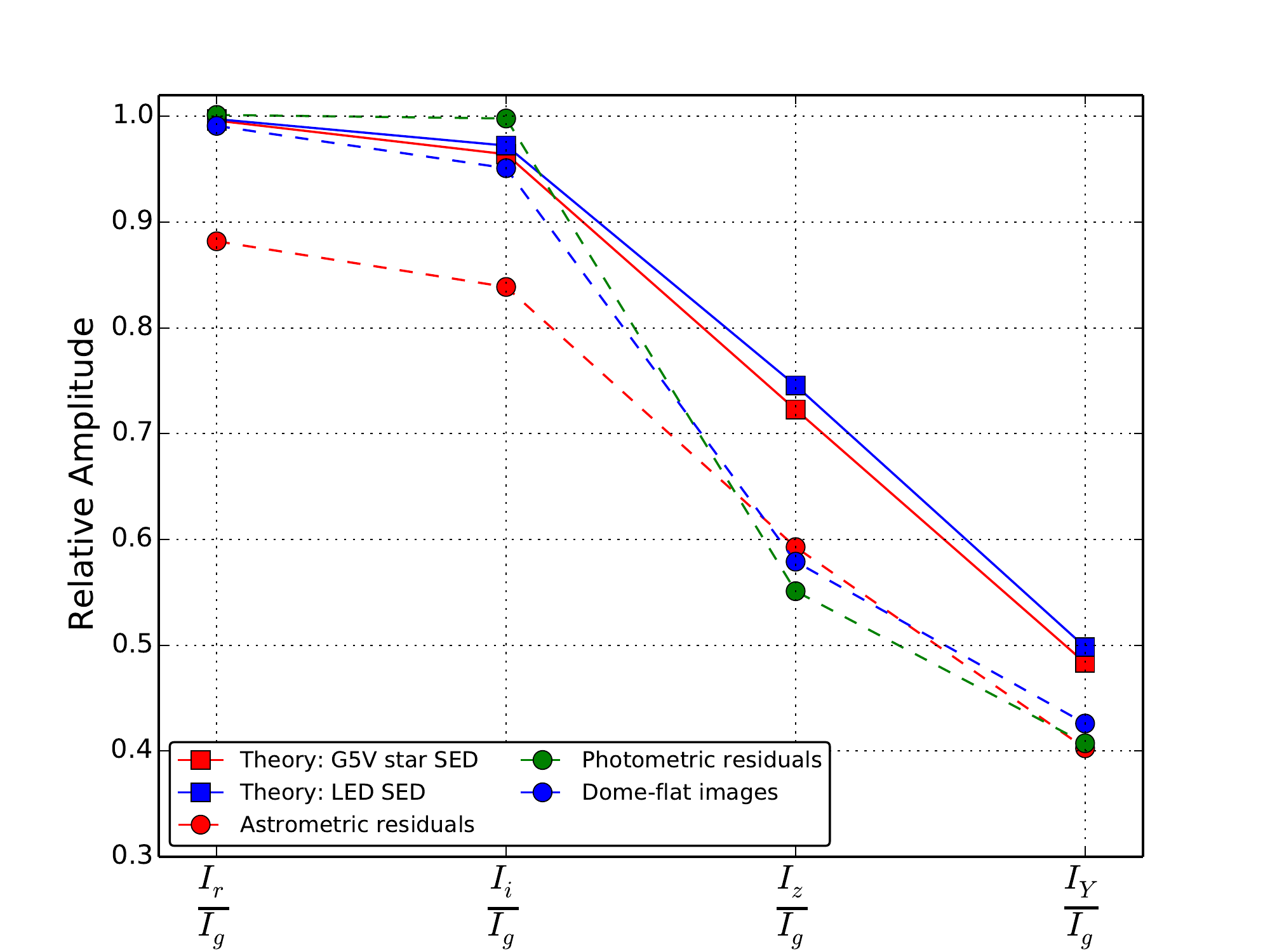}
\caption{Relative amplitude of the tree-ring profiles with respect to the $g$ band (\emph{i.e.}, the factor by which one must multiply the profile in the $g$ band to obtain the profiles in the other bands). The solid lines with squared symbols show the values estimated from Equation (\ref{amplitude}) by using the approximations described in the text. The solid, blue line uses the spectrum of a G5V-type star as source SED, whereas the red, solid line uses the spectrum of one of the high-power LEDs used to create the DECam dome flats, for each photometric band. The dashed lines with circular symbols show the average of the measured values over all DECam CCDs for amplifier A of each CCD (the values for amplifier B are similar): the red, blue, and green dashed lines are for the multiplicative factors obtained from the photometric residuals, the astrometric residuals (Figure \ref{plotfive}, upper panel), and the dome-flat images (Figure \ref{plotsix}, upper panel), respectively.}
\label{ploteight}
\end{figure}

Using Equation (\ref{delta_x}), the amplitude of the astrometric signatures in different bands as measured by using the star-flat data will be given by (setting $k\equiv b/a$):
\begin{equation}
 \Delta X_{\perp} (y) = \frac{A_{\perp}}{a} d\left \{ \frac{(y'/d)^2}{2k} - \frac{(y'/d)}{k^2} + \left( \frac{1}{k^3} - \frac{1}{k^2} \right) \ln \left( k \left( \frac{y'}{d}-1\right) +1 \right)  \right \} \bigg |_{y}^d 
\label{astrometric_shift}
\end{equation} 
 $\Delta X_{\perp}$ peaks at the origin (the back side of the device) and monotonically decreases to zero at the front gate ($y=d$), as expected. 

The amplitude of the ring profiles measured in the dome flats is given by the spatial derivative of Equation (\ref{astrometric_shift}) with respect to the transverse axis, $\partial_{x} \Delta X_{\perp}$.  For the particular case in which the transverse and electric fields are separable into functions that only depend on $x$ and $y$, the relative astrometric and photometric shifts as function of the substrate depth will be identical (the derivative will affect only the transverse component, and it will cancel out when amplitude ratios are calculated). 

\subsection{Measurements from dome- and star-flat images}
We used Equation (\ref{amplitude}) to calculate the relative amplitudes of the tree rings profiles in all bands, with respect to the $g$ band. For the absorption length of light in silicon, we used the values obtained from the model in \citet{rajkanan1979}, as calculated by D. Groom,\footnote{\url{http://snap.lbl.gov/ccdweb/ccd\_data.html}} for a temperature of $T=173$ K (the nominal working temperature of the DECam focal CCDs). The DECam flat-field images are created on a daily basis by illuminating a nearly Lambertian screen with 4 LEDs stations (with 7 hight-power LEDs each one) around the telescope top ring \citep{marshall2013}. We used the relative spectral energy distributions of the LEDs to make a comparison with the data from dome-flat images.\footnote{The spectral distributions of the LEDs are plotted in Figure 2 of \citet{marshall2013}. Each one of these relative SEDs is normalized to its maximum. The LEDs are chosen so one DES band is illuminated mostly by just one LED peaking at the central wavelength of the band. In our calculations, when two LEDs overlap in one band we chose one of them, as an approximation.} The data to produce the astrometric profiles of the tree rings (Figure \ref{plotfive}) were obtained by imaging several stars in different position of the focal plane. In this case we approximate the stars' SEDs by using a single spectral energy distribution of a G5V star \citep{pickles1998} in Equation (\ref{amplitude}). We calculate relative amplitudes between filters, so proportionality constants (\emph{e.g.}, $A_{\perp}$) will cancel out. To calculate $E_{\parallel}$, we use $V_{\text{sub}}=40$V, $V_{\text{J}}=-17$V, {$N_\text{D}=8.61\times10^{11}$ cm$^{-3}$ }, and $\epsilon_{\text{Si}}=11.7\epsilon_{0}$.  

We compare our calculations with the average of the measured relative amplitudes between the tree rings profiles in different bands in Figure \ref{ploteight}.  The measured values were obtained by using the tree-ring radial profiles as seen in the photometric residuals, the astrometric residuals, and the dome-flat images (see Figures \ref{plotthree}, \ref{plotfive}, and \ref{plotsix}).

Both the measured and calculated ratios diminish with wavelength, in agreement with the expectations from our electrostatic model. Also, the ratios from the dome-flat and photometric-residuals data trace each other, as was demonstrated in Figure \ref{plotthree} for the $i$ band. The discrepancies in the actual values could be due to the approximations performed to obtain expressions for the functions in Equation (\ref{amplitude}). For example, the astrometric shift $\Delta X_{\perp}$ depends explicitly on the star's SEDs. In addition, a more complex model for the $y$ dependence of the  transverse electric field might be needed.


\begin{thebibliography}{}

\bibitem[Abbott et al.(2005)]{abbott2005}
T. Abbott et al.,
\emph{The Dark Energy Survey}, (2005), arXiv:0510346

\bibitem[Albrecht et al.(2006)]{albrecht2006}
A. Albrecht et al., 
\emph{Report of the Dark Energy Task Force}, (2006), arXiv:0609591

\bibitem[Annis et al.(2010)]{annis2010}
J. Annis et al., 
\emph{The Dark Energy Survey science requirements document}, located at \url{http://des-docdb.fnal.gov:8080/cgi-bin/ShowDocument?docid=20\&version=32} (2010). Available on request. 

\bibitem[Altmannshofer et al.(2003)]{altmannshofer2003}
L. Altmannshofer et. al,
\emph{A material innovation for the electronic industry: floating zone single crystal silicon with 200mm diameter}, (2003), IEEE 15th Intl. Symp. on Power Semiconductor Devices and ICs Proc., Cambridge UK, 14th - 17th April 2003, pp. 325-328.

\bibitem[Antilogus et al.(2014)]{antilogus2014}
P. Antilogus et al., 
\emph{The brighter-fatter effect and pixel correlations in CCD sensors}, (2014), JINST 9 C03048, (arXiv:1402.0725)

\bibitem[Bertin \& Arnouts(1996)]{bertin1996}
E. Bertin \& S. Arnouts, 
\emph{SExtractor: Software for source extraction}, (1996), Astronomy \& Astrophysics Supplement 317, 393

\bibitem[Bertin(2006)]{bertin2006}
E. Bertin, 
\emph{Automatic Astrometric and Photometric Calibration with SCAMP} (2006), ASP Conference Series, Vol. 351, 2006, C. Gabriel, C. Arviset, D. Ponz, and E. Solano, eds., p. 112

\bibitem[Diehl et al.(2008)]{diehl2008}
H. T. Diehl et al.,  
\emph{Characterization of DECam Focal Plane Detectors}, (2008),  Proc. SPIE Int.Soc.Opt.Eng. 7021 

\bibitem[Diehl(2012)] {diehl2012}
H. T. Diehl, 
\emph{The Dark Energy Survey Camera (DECam)}, (2012), Physics Procedia, Proceedings of the 2nd International Conference on Technology and Instrumentation in Particle Physics (TIPP 2011), V37 pp 1332-1340.

\bibitem[Diehl et al.(2014)]{diehl2014}
H. T. Diehl et al., 
\emph{The Dark Energy Survey and Operations: Year 1}, (2014),  Proc. SPIE 9149, 9149-31 (in press).

\bibitem[Derylo et al.(2006)]{derylo2006}
G. Derylo et al., 
\emph{0.25mm-Thick CCD Packaging for the Dark Energy Survey Camera Array} (2006), 
Proc. SPIE, Volume 6276.

\bibitem[Desai et al.(2012)]{desai2012}
S. Desai et al., 
\emph{The Blanco Cosmology Survey: data acquisition, processing, calibration, quality diagnostics, and data release}, (2012), The Astrophysical Journal, Volume 757, Issue 1, article id. 83, 22 pp. 

\bibitem[Estrada et al.(2010)]{estrada2010}
J. Estrada et al., 
\emph{Focal plane detectors for the Dark Energy Survey}, (2010), 
Ground-based and Airborne Instrumentation for Astronomy III. Edited by McLean, Ian S.; Ramsay, Suzanne K.; Takami, Hideki. Proceedings of the SPIE, Volume 7735, article id. 77351R, 11 pp. 

\bibitem[Fairfield et al.(2006)]{fairfield2006}
J. Fairfield et al., 
\emph{Reduced charge diffusion in thick, fully depleted CCDs with enhanced red sensitivity}, (2006), IEEE Transactions on Nuclear Science, vol. 53, no. 6.

\bibitem[Flaugher et al.(2012)]{flaugher2012}
B. Flaugher et al., 
\emph{Status of the Dark Energy Survey Camera (DECam) project}, (2012),  Proc. SPIE 8446, Ground-based and Airborne Instrumentation for Astronomy IV, 844611.

\bibitem[Holland et al.(2003)]{holland2003}
S. Holland et al., 
\emph{Fully depleted, back-illuminated charge-coupled devices fabricated on high-resistivity Silicon}, (2003),  IEEE Transactions on Electron Devices, vol. 50, no. 1.

\bibitem[Holland et al.(2007)]{holland2007}
S. Holland et al., 
\emph{Fabrication of back-illuminated, fully depleted charge-coupled devices}, (2007),  Nucl. Instrum. Methods, A579, 653-657.

\bibitem[Holland et al.(2009)]{holland2009}
S. Holland et al.,
\emph{Device design for a 12.3-megapixel, fully-depleted, back-illuminated, high-voltage compatible charge-coupled device}, (2009), IEEE Transactions on Electron Devices, vol. 56, no. 11.

\bibitem[Holland et al.(2014)]{holland2014}
S. Holland et al., 
\emph{Physics of Fully Depleted CCDs}, (2014), JINST 9 C03057, (arXiv:1403.6185)

\bibitem[Ivezic et al.(2008)]{ivezic2008}
Z. Ivezic et al., 
\emph{LSST: from Science Drivers to Reference Design and Anticipated Data Products}, (2008), arXiv:0805.2366

\bibitem[Jarvis(2014)]{jarvis2014}
M. Jarvis, 
\emph{Challenges for precision shape measurements}, (2014), JINST 9 C03017

\bibitem[Kaiser et al.(2010)]{kaiser2010} 
N. Kaiser et al.,
\emph{The Pan-STARRS wide-field optical/NIR imaging survey}, (2010),
Proc. of SPIE, 7733, 77330E-1.

\bibitem[Komiyama et al.(2010)]{komiyama2010} 
Y. Komiyama et al.,
\emph{Hyper Suprime-Cam: Camera Design},  (2010)
Proc. of SPIE, 7735, 77353F-1.

\bibitem[Kotov et al.(2006)]{kotov2006}
I. Kotov et al., 
\emph{Electric fields in nonhomogeneously doped silicon. Summary of simulations.}, (2006),  Nuclear Instruments and Methods in Physics Research A,  568, 41-45

\bibitem[Kotov et al.(2011)]{kotov2011}
I. Kotov et al., 
\emph{Study of pixel area variations in fully depleted thick CCD},  (2010), 
Proc. SPIE 7742, High Energy, Optical, and Infrared Detectors for Astronomy IV, 774206; doi:10.1117/12.856519

\bibitem[Kuhlman et al.(2011)]{kuhlman2011}
S. Kuhlmann et al., 
\emph {Narrow-beam X-ray tests of CCD edge response}, (2011), Experimental Astronomy, 29, p135-144

\bibitem[Lupton(2014)]{lupton2014}
R. Lupton, 
\emph{Consequences of thick CCDs on image processing}, (2014),  JINST 9 C04023

\bibitem[Manfroid(1995)]{manfroid1995}
J. Manfroid,
\emph{On CCD standard stars and flat-field calibration}, (1995), A\&AS, 113, 587 

\bibitem[Marshall et al.(2013)]{marshall2013}
J. L. Marshall et al.,
\emph{DECal: A Spectrophotometric Calibration System for DECam}, (2013), arXiv:1302.5720

\bibitem[Mohr et al.(2008)]{mohr2008}
J. Mohr et al., 
\emph{The Dark Energy Survey data management system}, (2008), Observatory Operations: Strategies, Processes, and Systems II. Edited by Brissenden, Roger J.; Silva, David R. Proceedings of the SPIE, Volume 7016, article id. 70160L, 16 pp.

\bibitem[O'Connor(2014)]{oconnor2014}
P. O'Connor, 
\emph{Spot Scan Probe of Lateral Field Effects in a Thick Fully-Depleted CCD}, (2014), JINST 9 C03033, ( arXiv:1402.5454)

\bibitem[Pickles(1998)]{pickles1998}
A. Pickles, 
\emph{A Stellar Spectral Flux Library: 1150-25000 A}, (1998), Publications of the Astronomical Society of the Pacific, 110, p863

\bibitem [Plazas \& Bernstein(2012)]{plazas2012}
A. A. Plazas \& G. Bernstein, 
\emph{Atmospheric dispersion effects in weak lensing measurements} (2012), Publications of the Astronomical Society of the Pacific, vol. 124, No. 920, pp 1113-1123

\bibitem[Rajkanan et al.(1979)]{rajkanan1979}
K. Rajkanan et al.,
\emph{Absorption coefficient of silicon for solar cell calculations} (1979), Solid-State Electronics, vol. 22, p. 793-795.

\bibitem[Rasmussen(2014)]{rasmussen2014}
A. Rasmussen, 
\emph{Pixel area variations in sensors: a novel framework for predicting pixel fidelity and distortion in flat field response} (2014), JINST 9 C04027, (arXiv:1403.3317)

\bibitem[Smith \& Rahmer(2008)]{smith2008}
R. Smith \& G. Rahmer, 
\emph{Pixel area variation in CCDs and implications for precise photometric calibration} (2008),
SPIE Conference Series, 7021

\bibitem[Stubbs(2014)]{stubbs2014}
C. Stubbs, 
\emph{Precision astronomy with imperfect fully depleted CCDs: an introduction and a suggested lexicon}, (2014), JINST 9 C03032, (arXiv:1312.2313)

\bibitem[Tucker et al.(2007)]{tucker2007}
D. Tucker et al., 
\emph{The photometric calibration of the Dark Energy Survey}, (2007), 
The Future of Photometric, Spectrophotometric and Polarimetric Standardization, ASP Conference Series, Vol. 364, Proceedings of a conference held 8-11 May, 2006 in Blankenberge, Belgium. Edited by C. Sterken. San Francisco: Astronomical Society of the Pacific, 2007., p.187

\bibitem[von Ammon \& Herzer(1984)]{vonAmmon1984}
W. von Ammon \& H. Herzer, 
\emph{The Production and Availability of High Resistivity Silicon for Detector Application}, (1984), Nuclear Instruments and Methods in Physics Research v. 226, 94-102

\bibitem[Weinberg et al.(2013)]{weinberg2013}
D. Weinberg et al., 
\emph{Observational probes of cosmic acceleration}, (2013), Physics Reports, Volume 530, Issue 2, p. 87-255

\end{thebibliography}
\end{document}